\documentclass[11pt, a4paper, logo, copyright, nonumbering]{googledeepmind}

\usepackage[authoryear, sort&compress, round]{natbib}
\usepackage{hhline} 
\usepackage{highlight}
\usepackage{xcolor}
\usepackage{hyperref}
\renewcommand{\opacity}{50}

\title{Consensus, dissensus and synergy between clinicians and specialist foundation models in radiology report generation}

\correspondingauthor{$\{\text{rtanno},\text{barrettdavid},\text{alankarthi},\text{iraktena}\}$@google.com}


\reportnumber{001} 

\renewcommand{\today}{}

\author[1,*]{Ryutaro Tanno}
\author[1,*]{David G.T. Barrett}
\author[2]{Andrew Sellergren}
\author[1]{Sumedh Ghaisas}
\author[1]{Sumanth Dathathri}
\author[1]{\\ Abigail See}
\author[1]{Johannes Welbl}
\author[2]{Karan Singhal}
\author[1]{Shekoofeh Azizi}
\author[2]{Tao Tu}
\author[2]{Mike Schaekermann}
\author[1]{\\ Rhys May}
\author[2]{Roy Lee} 
\author[2]{SiWai Man}
\author[1]{Zahra Ahmed}
\author[1]{Sara Mahdavi}
\author[2]{Yossi Matias}
\author[1]{Joelle Barral}
\author[1]{\\Ali Eslami}
\author[1]{Danielle Belgrave}
\author[2]{Vivek Natarajan}
\author[2]{Shravya Shetty}
\author[1]{Pushmeet Kohli}
\author[1]{Po-Sen Huang}
\author[2]{\\Alan Karthikesalingam}
\author[1,*]{Ira Ktena}

\affil[1]{Google DeepMind}
\affil[2]{Google Research}
\affil[*]{Equal technical contributions}

\begin{abstract}
Radiology reports are an instrumental part of modern medicine, informing key clinical decisions such as diagnosis and treatment. The worldwide shortage of radiologists, however, restricts access to expert care and imposes heavy workloads, contributing to avoidable errors and delays in report delivery. While recent progress in automated report generation with vision-language models offer clear potential in ameliorating the situation, the path to real-world adoption has been stymied by the challenge of evaluating the clinical quality of AI-generated reports. In this study, we build a state-of-the-art report generation system for chest radiographs, \textit{Flamingo-CXR}, by fine-tuning a well-known vision-language foundation model on radiology data. To evaluate the quality of the AI-generated reports, a group of 16 certified radiologists provide detailed evaluations of AI-generated and human written reports for chest X-rays from an intensive care setting in the United States and an inpatient setting in India.  At least one radiologist (out of two per case) preferred the AI report to the ground truth report in over 60$\%$ of cases for both datasets. Amongst the subset of AI-generated reports that contain errors, the most frequently cited reasons were related to the location and finding, whereas for human written reports, most mistakes were related to severity and finding.  This disparity suggested potential complementarity between our AI system and human experts, prompting us to develop an assistive scenario in which \textit{Flamingo-CXR} generates a first-draft report, which is subsequently revised by a clinician. This is the first demonstration of clinician-AI collaboration for report writing, and the resultant reports are assessed to be equivalent or preferred by at least one radiologist to reports written by experts alone in 80$\%$ of in-patient cases and 66$\%$ of intensive care cases.
\end{abstract}

\begin{document}

\maketitle

\section{Introduction}

Radiology plays an integral and ever increasing function in modern medicine by informing diagnosis, treatment and management of patients through medical imaging. However, the present global shortage of radiologists limits access to expert care and necessitates heavy workloads, resulting in undesirable delays and errors in clinical decisions~\citep{maru2010turning,rimmer2017radiologist}. In the last decade, we have witnessed a remarkable promise of Artificial Intelligence (AI) algorithms as assistive technology for improving the access, efficiency and quality of radiological care, with more than 200 FDA-approved commercial products developed by companies based in more than 20 countries~\citep{rajpurkar2023current} and approximately one in every three radiologists in the US already benefiting from AI as part of their clinical workflow~\citep{allen20212020}.

The vast majority of these approved AI applications, however, only focus on classification and quantification of very specific pathologies \citep{milam2023current}. In practice, clinical radiology is much more than an accumulation of such narrow interpretive tasks, as findings must be communicated with appropriate nuance, synthesized in a broader clinical context and combined with an overall impression and recommendations that are useful to patient care. Radiologist experts use natural language to communicate this synthesis of the imaging findings, their overall impression and recommendations in the form of written reports. The recent progress in AI for modelling vision and language data simultaneously \citep{baltruvsaitis2018multimodal,guo2019multimodalsurvey,alayrac2022flamingo,li2023multimodal}, coupled with the growing availability of digitized multi-modal radiology data, has enabled the possibility of developing an automatic report generation system that is capable of producing a complete free-text description of the medical image \citep{chen2020,krishnan2021,miura2021,nicolson2023,yan2023style}. Framing report generation as the key north star for a useful radiology AI system matches more closely how radiologists influence care in practice, and allows for a more fine-grained and diverse description of the relevant findings that can be tailored to the needs of a given clinical scenario, including aspects such as location, size and severity, ambiguity, relation to clinical context of specific pathologies or their impact on onward care and more \citep{bannur2023learning}.

Despite the increasing number of publications on AI-based report generation and its potential in improving the radiology workflow, automated report generation has not yet been widely adopted in real practice \citep{milam2023current}. Several unmet needs represent key barriers to automated reporting achieving real-world impact. One notable obstacle is the difficulty of meaningfully evaluating the clinical quality of generated reports. The high degree of freedom in free-form reports introduces a wide range of possible errors to measure and phenotype. Exacerbating this, the desirable contents of a report differ between clinical settings (e.g., an emergency setting vs a medical check-up), geographic regions \citep{hartung2020create} and preferred approaches to standardization \citep{kahn2009toward}. Prior works have approached this challenge by proposing automated metrics for evaluating the clinical quality of generated reports  \citep{liu2019clinically,jain2021radgraph,khanna2023radgraph2,yu2022evaluating} but significant limitations remain. Firstly, there has been a paucity of comprehensive evaluation of automated reports against reports produced by human experts (certified radiologists), which are known to themselves vary in quality. Despite impressive progress in automated metrics for report quality, only one study~\citep{tu2023generalist} has directly assessed whether AI-generated reports were considered preferable to those by human experts while only~\citet{huang2023generative} have evaluated their utility in practice in a narrow clinical setting. At the same time, there is a lack of detailed analysis of the sources of such preference. Secondly, prior work has only evaluated AI-generated reports as stand-alone artefacts, meaning the utility of these systems as assistive tools remains unknown. Evaluation in such clinician-AI collaboration scenarios is arguably more realistic, given that most AI tools that have been approved for clinical decision-making play an assistive rather than autonomous role in care delivery \citep{harvey2020fda,norden2022ai}. 

\begin{figure}[t!]
    \centering
    \includegraphics[width=\linewidth]{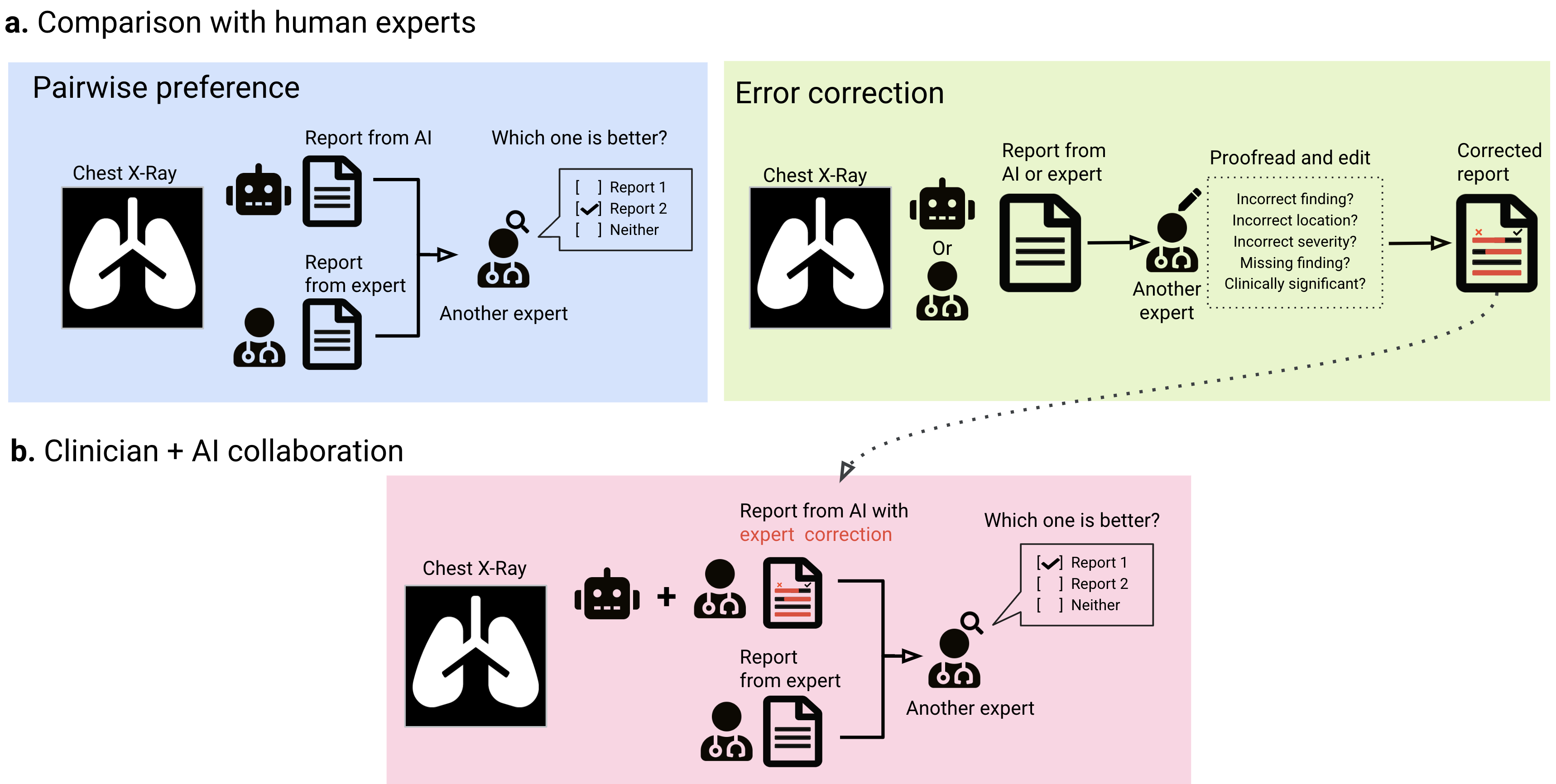}
    \caption{\small \textbf{Schematic overview of our human evaluation framework}. \textbf{a.} To measure the gaps between an AI model and human experts on the stand-alone report writing task, we devise two evaluation schemes: i) a \textit{pairwise preference} test where a certified expert is given two reports without knowing their sources (one from a model and the other from a radiologist) and assess their relative usefulness; ii) an \textit{error correction} task where a single report (from either a model or an expert) is evaluated carefully and edited if required. For the second task, the expert is further asked to detail the reason for each correction and whether the error is clinically significant. \textbf{b.} We further measure the utility of the AI-based report generation system in an assistive scenario where the AI model first generates a report and the human expert revises as needed. For this task, we repeat the same pairwise preference test in \textbf{a.}(left) but this time, the expert is asked to compare an AI-generated report corrected with human edits against a report written by human alone. We perform this evaluation on two datasets, one acquired in India and another in the US.}

    \label{fig:overview}
\end{figure}
In addition to the above evaluation challenges, there remains considerable headroom for improvement in clinical accuracy of existing AI report generation models \citep{yu2022evaluating}. Recent breakthroughs in multi-modal foundation models \citep{li2023multimodal} have demonstrated that AI systems trained on a vast quantity of unlabelled data can be adapted and achieve state-of-the-art accuracy in a wide range of downstream specialised tasks, including biomedical problems \citep{li2023llavamed}. However, most existing report generation models \citep{chen2020,krishnan2021,miura2021,nicolson2023} are built from scratch, neglecting the likely useful transfer of knowledge from such pre-trained models. By leveraging advances accrued through large-scale pretraining of vision-language models and tailoring them to a specific medical task, there is an opportunity to build an even more powerful report generation system.

In this work, we directly address these key unmet needs for AI report generation. We present Flamingo-CXR, a system for AI report generation predicated on a recent vision-language foundation model \citep{alayrac2022flamingo} that achieves state-of-art performance in multiple automated metrics. We evaluate Flamingo-CXR in more than one clinical context and geography -- both intensive care in the United States, and inpatient care delivery in India and move beyond automated metrics to a detailed human evaluation of the reports generated by this system, including a direct comparison of clinicians' preferences for AI reports versus human reports. We evaluate the system in an autonomous as well as assistive context. Figure \ref{fig:overview} shows an overview of the proposed evaluation framework. Taken together, our contributions outline a system with potential clinical applicability while more precisely defining areas in which performance should be improved. 

\section{Results}

\subsection{Adapting a vision-language foundation model to report generation}

We developed and evaluated our automatic report generation model on two large datasets of chest X-ray (CXR) images and corresponding radiology reports from the USA and India. Chest radiography offers a valuable testbed for automatic report generation systems as the most widely used thoracic imaging modality in the world \citep{nabulsi2021deep}. Even for such a specific domain the contents of radiology reports differ widely between geographic regions and clinical contexts. To account for such variations, we employed the combination of the \textit{MIMIC-CXR} dataset \citep{johnson2019mimic, johnson2019mimicdatabase, goldberger2000physiobank}, the largest public CXR dataset, acquired in the emergency department of the Beth Israel Deaconess Medical Center in the USA, and another private research dataset of a similar scale, which we refer to as \textit{IND1} \citep{nabulsi2021deep}, obtained from five regional centers across a large hospital group in India (Bangalore, Bhubaneswar, Chennai, Hyderabad, and New Delhi). Appendix~\autoref{sec:methods_datasets} provides details of these datasets.

Our report generation model is built by fine-tuning \textit{Flamingo}~\citep{alayrac2022flamingo} -- a state-of-the-art vision-language foundation model with impressive performance on data-efficient adaptation to new tasks -- on the radiology report generation task, with an effective combination of modern regularisation and adaptation techniques. Flamingo has a flexible transformer-based multi-modal sequence-to-sequence architecture that allows for integrating a mixture of medical images and reports without any model modifications (see \autoref{subsec:methods_model} for the details of the architecture, optimisation and inference). Our model is trained to generate both the `\textit{findings}' and `\textit{impression}' sections of the report for a frontal view (anterior-posterior or posterior-anterior) of the chest radiograph, which typically captures all the relevant observations the radiologist makes in a study.

\subsubsection*{Automated measures of report quality}
 We report performance on established automated metrics in order to facilitate comparison with prior studies, considering two groups of metrics. The first category is the \textit{natural language generation (NLG) metrics} that include scores such as CIDEr \citep{Ramakrishna2015}, BLEU score \citep{papineni-etal-2002-bleu} and Rouge-L \citep{lin2004rouge,Ramakrishna2015}, which are commonly-quoted measures of report quality. However, multiple studies \citep{liu2019clinically,pmlr-v116-boag20a,maynez2020faithfulness} have recently highlighted the inadequacy of these NLG metrics in assessing factual correctness and consistency, key properties determining the clinical utility and quality of radiology reports. 
 We further compute another set of metrics that are specifically designed to measure the accuracy of descriptions for relevant clinical findings, and we refer to them as \textit{clinical metrics}. Specifically, following the prior works~\citep{liu2019clinically,chen2020,miura2021,bannur2023learning}, we report the micro-average F1 score across 14 distinct categories (\texttt{atelectasis}, \texttt{cardiomegaly}, \texttt{consolidation}, \texttt{edema}, \texttt{enlarged cardiomediastinum}, \texttt{fracture}, \texttt{lung lesion}, \texttt{lung opacity}, \texttt{no finding}, \texttt{pleural effusion}, \texttt{pleural other}, \texttt{pneumonia}, \linebreak \texttt{pneumothorax} and \texttt{support devices}) related to thoracic diseases and support devices. To ensure a fair comparison with prior publications on the MIMIC-CXR dataset, we employ a previously-published labelling software, namely CheXpert~\citep{irvin2019chexpert}, to extract labels of findings from the reports automatically. For the IND1 dataset, published results on classification performance are unavailable, so we instead employ the labels of these findings collected in a separate study \citep{ahn2022association} from a group of $18$ certified radiologists in the USA, and use the corresponding consensus labels as ground-truths. This way, we aim to mitigate the known inaccuracy of the CheXpert labeller software and have a test set with a more reliable metric of clinical factual correctness. Finally, to align with more recent studies \citep{tu2023generalist,yu2022evaluating}, we also report the \textit{Radgraph} score \citep{jain2021radgraph,khanna2023radgraph2}, which additionally accounts for not only the presence of these findings but also the relationships between them and other image features (e.g., anatomical locations). All the results are reported on held-out test data that were not used to train or tune the model.

\begin{table}[t]
    \centering
    \caption{\small \textbf{Comparison of automatic report generation metrics on the MIMIC-CXR dataset.} The column `Sections' indicates which sections of the radiology reports are generated by the respective models. Note that the metrics are retrieved from the corresponding publications. CheXpert F1 (all) denotes the micro-averaged F1 score across all 14 categories of findings while CheXpert F1 (top5) shows the same metric but over the most prevalent 5 categories (namely, \texttt{atelectasis}, \texttt{cardiomegaly}, \texttt{edema},
    \texttt{consolidation} and \texttt{pleural effusion}) in the MIMIC-CXR dataset. For all metrics, the higher the better, and the best results are shown in bold. An extended version with NLG metrics are provided in Appendix Table~\ref{tab:table_mimic_appendix}.}
    \label{tab:table_mimic}
    \scriptsize
    \centering
    \begin{tabular}{|*{5}{c|}}
        \hhline{~|~*{3}{|-}}
        \multicolumn{1}{c}{} & \multicolumn{1}{c|}{} & \multicolumn{3}{c|}{\textbf{Clinical Metrics}}  \\\hline
        \textbf{Model} & \textbf{Sections} & \textbf{CheXpert F1 (all)} & \textbf{CheXpert F1 (top 5)} & \textbf{Radgraph F1}\\ \hline
        CXR-RePaiR \citep{krishnan2021} & Findings only & 0.281 & - & 0.091   \\ \hline 
        $\mathcal{M}^2$ Transformer \citep{miura2021} & Findings only  & - & 0.567 & 0.220  \\ \hline
        RGRG \citep{tanida2023interactive} & Findings only &  0.447 & 0.547 & -   \\ \hline
        METransformer \citep{wang2023metransformer} & Findings only &  0.311 & - & -   \\ \hline
        Med-PaLM-M, 12B \citep{tu2023generalist}& Findings only & 0.514 & 0.565 & \textbf{\textit{0.252}}   \\ \hline \hline
        R2Gen \citep{chen2020} &  Findings + Impressions &  0.228 & 0.346 & 0.134   \\ \hline
         WCT \citep{yan2021weakly} &  Findings + Impressions &  0.294 & - & 0.143   \\ \hline
         CvT-21DistillGPT2 \citep{nicolson2023} &  Findings + Impressions & 0.384 & - & 0.154  \\ \hline
        BioVil-T \citep{bannur2023learning} &  Findings + Impressions & 0.317 & - & -   \\ \hline
        R2GenGPT \citep{wang2023r2gengpt} &  Findings + Impressions & 0.389 & - & -   \\ \hline 
        \cellcolor{blue!15} Flamingo-CXR (Ours) &  Findings + Impressions & \textbf{0.519} & \textbf{0.580} & 0.205  \\ \hline
    \end{tabular}
\end{table}

Table~\ref{tab:table_mimic} shows a performance comparison between the state-of-the-art methods and our model (\textit{Flamingo-CXR}) on the MIMIC-CXR dataset through the lens of clinical metrics. In terms of the overall F1 score, our model achieves competitive performance against the prior methods, marking $1\%$ and $15\%$ relative increases to the score of 0.519 from the second best result of 0.514 \citep{tu2023generalist} and the third best result of 0.447 \citep{tanida2023interactive}, both of which were published in the same year as this work. Furthermore, amongst the methods capable of generating both the `findings' and `impression' sections of the report (as denoted in the Sections column), we have outperformed the current SoTA method (CvT-21DitillGPT2 \citep{nicolson2023}) by a large margin, attaining $33 \%$ improvement from $0.154$ to $0.205$. While Med-PaLM-M \citep{tu2023generalist} and $\mathcal{M}^2$-Transformer report higher scores, we highlight that these methods only generate the `findings' sections of the reports and as such, a direct comparison in terms of a graph-based metric such as Radgraph F1 is challenging. In terms of the NLG metrics (CIDEr, BLEU4 and Rouge), the results are mixed; we achieve competitive BLEU4 and Rouge scores while attaining a compromised CIDEr score (see~\autoref{tab:table_mimic_appendix}). This is also consistent with the established observation that NLG metrics do not reflect the clinical accuracy of the generated reports \citep{liu2019clinically,pmlr-v116-boag20a,yu2022evaluating}, for which our model, in particular, confers an improvement over the relevant prior methods.

\subsubsection*{Disease classification in comparison to human radiologists}

\begin{figure}[t!]
    \centering
    \includegraphics[width=\linewidth]{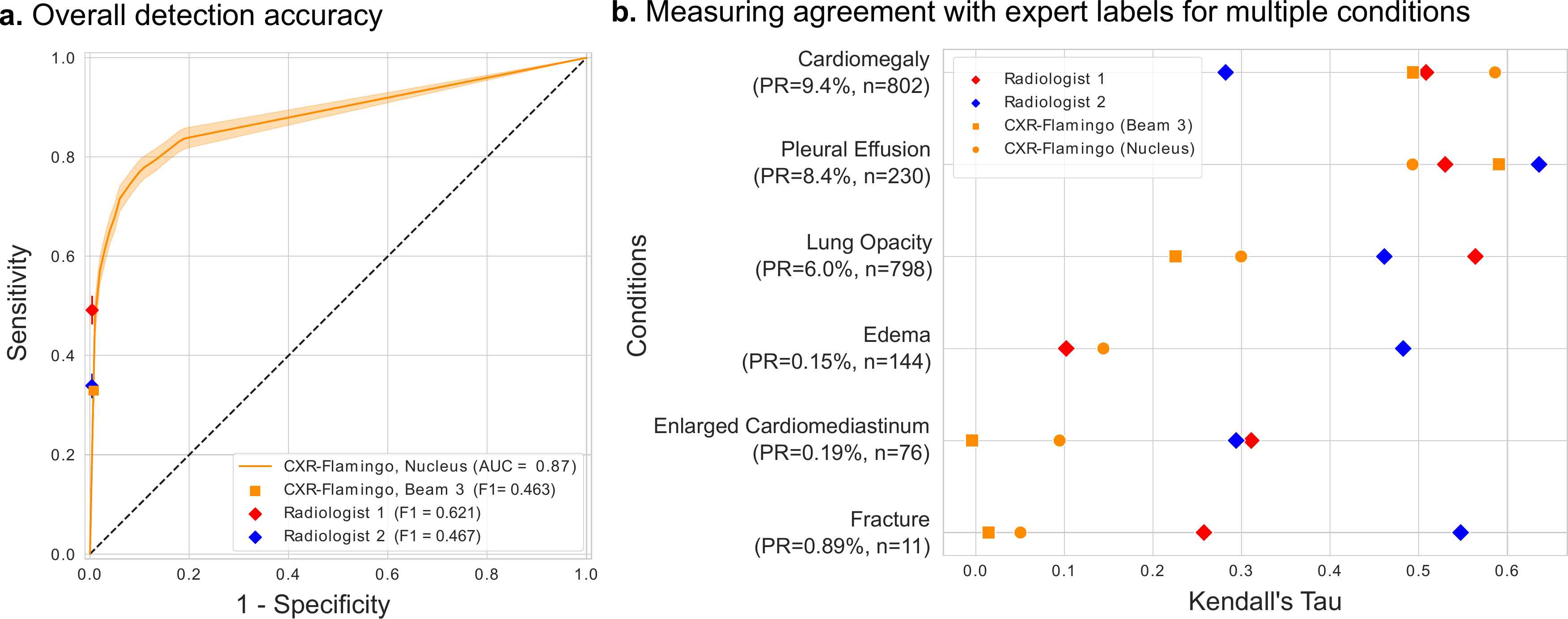}
    \caption{\small \textbf{Comparison of detection accuracy with expert labels on the IND1 dataset.} For both figures, the ground truth labels are defined as the majority vote among the 5 labels obtained from the pool of 18 certified radiologists. \textbf{a.} The receiver operating characteristic (ROC) curve of the Flamingo-CXR report generation model with stochastic generation method (Nucleus), shown along with the sensitivity and 1 - specificity pairs for two certified radiologists. The operating point of our model with the default deterministic inference scheme (Beam 3) is also shown. The details of the two inference algorithms are available in Appendix~\autoref{subsec:methods_model}. The curve and the metrics are micro-averaged across 6 conditions (cardiomegaly, pleural effusion, lung opacity, edema, enlarged cardiomediastinum and fracture) for which the labels were collected. \textbf{b.} Kendall's tau coefficients with respect to the expert labels are shown for the two held-out radiologists as well as for two inference schemes of our Flamingo-CXR model. Here we use the `soft' labels derived by averaging over the available annotations instead of the majority vote labels as the target for computing the metric. On the vertical axis, the prevalence rates (PRs) of the respective conditions in the training set and their sample size in the test set are also shown. Here the target labels are the probabilities over the presence of the respective conditions calculated by averaging the binary condition labels from the expert pool. 
    }
    \label{fig:apollo_overall}
\end{figure}

For the IND1 dataset, \autoref{fig:apollo_overall}.\textbf{a} shows that the generated reports of our model are overall as accurate (in terms of the micro-averaged F1 score) as one of the two radiologists in describing six clinical conditions in chest radiographs (namely, \texttt{cardiomegaly}, \texttt{pleural effusion}, \texttt{lung opacity}, \texttt{edema}, \texttt{enlarged cardiomediastinum} and \texttt{fracture}). We highlight that the ground truth labels here are derived from the majority votes of 5 annotations per example acquired by a separate group of 18 experts and, thus, should provide more reliable labels than the ones extracted from the CheXpert labeller \citep{irvin2019chexpert} (which was used for the MIMIC-CXR dataset). On the other hand, to generate the binary labels from the generated reports from Flamingo-CXR, the CheXpert labeller is used as before. For conditions that are frequent in the training dataset such as cardiomegaly and pleural effusion, we attain comparable or even superior agreement with the experts labels (as measured in the Kendall's tau coefficients) with respect to the two held-out radiologists (\autoref{fig:apollo_overall}.\textbf{b}). On the other hand, for under-represented conditions such as \texttt{edema} and \texttt{enlarged cardiomediastinum} with extremely low prevalence rates ($0.19\%$ and $0.15\%$, respectively), the agreement scores of our model are lower than the two radiologists. The ROC curves for the individual conditions (Supplementary Figure~\ref{fig:apollo_roc_per_condition}) exhibit patterns consistent with such variation in the accuracy across conditions of different prevalence. We also report performance using the consistent set of NLG and clinical metrics for completeness in Supplementary Table~\ref{tab:table_apollo}.

\subsection{Expert evaluation of AI-generated and human-written reports}
\label{subsec:expert_evaluation}

Accumulated evidence has shown that automatic report generation metrics fail to appropriately evaluate many nuanced issues of radiology reports \citep{yu2022evaluating}. To achieve a more fine-grained and realistic assessment of the clinical quality of radiology reports generated by our model, we conduct an expert evaluation for reports in both the MIMIC-CXR and IND1 datasets. Moreover, in order to document human errors in report writing and to characterise differences in quality with our AI system, we also evaluate the original reports (that we have treated as ground-truths) by obtaining additional readings from different radiologists than the ones who provided the original reports.

We recruit a group of $16$ certified radiologists in India to perform two complementary evaluation tasks, namely (1)~\textit{pairwise preference test} and (2) an~\textit{error correction} task. Figure~\ref{fig:overview}.\textbf{a} illustrates an overview of these two evaluation tasks. We ensure to have each report evaluated by two radiologists to measure inter-rater variability. 

In total, 554 cases were evaluated by expert radiologists in the two tasks: 32 normal and 272 abnormal cases from the \textit{MIMIC-CXR} dataset, and 50 normal and 200 abnormal cases from the \textit{IND1} dataset. We ensure coverage of multiple abnormal cases for both datasets, as we found classification quality to vary significantly across conditions. Section~\ref{sec:human_eval_details} provides additional details on the sample selection logic. 

\subsubsection*{Pairwise preference test}
In this evaluation task, radiologists are provided with (i) a frontal view of a CXR image, (ii) a radiology report generated by our AI system and (iii) the original report written by a radiologist, and are asked to assess the relative usefulness of the two reports for the given image. For each case, the raters are unaware of which report is the original and which one is generated by our model, and are requested to describe their preference out of three options; report A, report B, or equivalence between the two (i.e., ``neither is better than the other''). Furthermore, they are asked to provide a justification for their preference to better understand strengths and limitations of either.~\autoref{fig:interface_pairwise_preference} illustrates the labelling interface used by the radiologists for this evaluation task. We note that the assignment of the original and the generated reports to option A and B is completely random. 

\begin{figure}[t!]
    \centering
    \includegraphics[width=\linewidth]{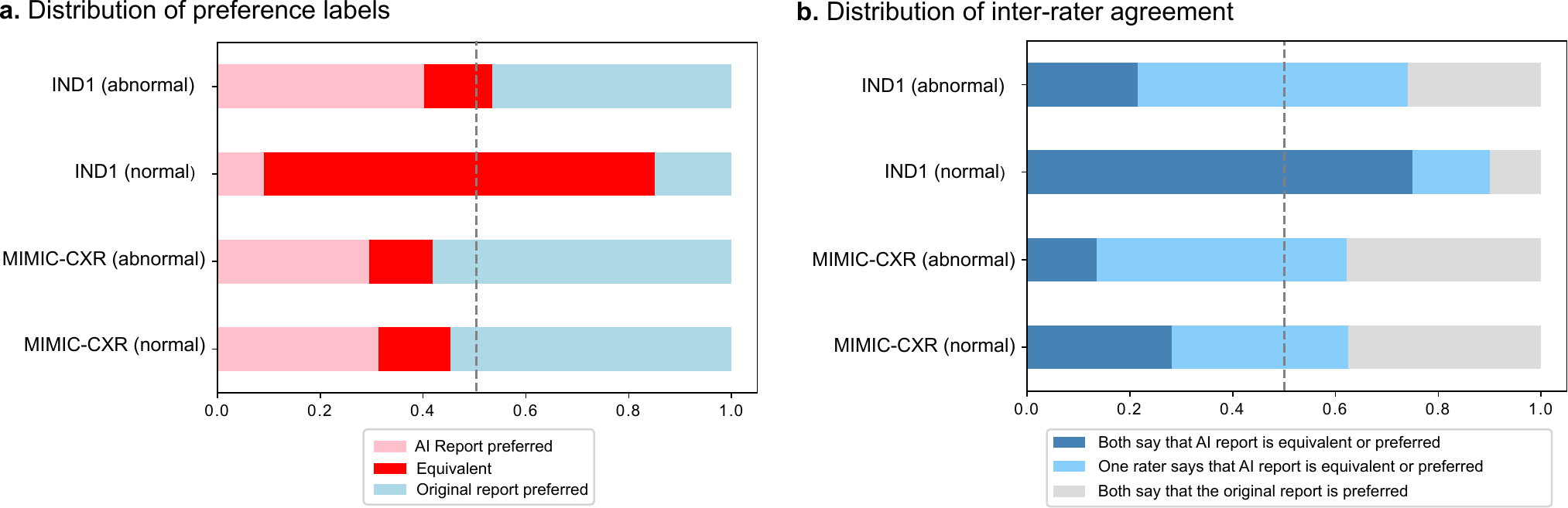}
    \caption{\small \textbf{Results of pairwise preference test}. \textbf{a.} Overall percentage of cases for which the AI report was preferred (pink), the original report was preferred (light blue) or the two were considered of equivalent quality (red). It is worth noting that inter-rater variability in their preferences is not captured in this plot. \textbf{b.} Inter-rater agreement in their assessment that the Flamingo-CXR report is preferred or of equivalent quality to the ground truth report (dark blue) or that the ground truth report is preferred (grey). Each case is annotated by two radiologists. Light blue corresponds to cases for which at least one rater did not prefer the ground truth report. Across both normal and abnormal subsets of the IND1 and MIMIC-CXR datasets, the AI report was considered better or equivalent to the original report in at least 2 out of 3 cases.} 
    \label{fig:pairwise_preference}
\end{figure}

\autoref{fig:pairwise_preference} summarises the preferences of expert raters in this task.~\autoref{fig:pairwise_preference}.\textbf{a} illustrates the percentage of instances with a preference for the original or AI report, divided into normal and abnormal cases. Across both datasets, generated reports by the AI system were often considered preferable or equivalent to the ground truth report. For IND1, in particular, generated reports were considered at least equivalent (or better) for more than 50\% of the abnormal cases and 85\% of the normal cases. It becomes apparent that MIMIC-CXR reports are more challenging to model, as is evident by the lower preference rates for the AI generated reports in comparison to IND1. This is likely due to the diversity of the original reports in this dataset. In order to better understand the inter-rater concordance, we further plot the percentage of cases with at least one rater considering the AI report equivalent or better. Across both datasets, the AI reports were considered at least equivalent (or better) in comparison to the ground truth reports by at least one rater in more than 60\% of cases (\autoref{fig:pairwise_preference}\textbf{.b}). For the normal cases, in particular, this held true for 90\% of cases. The largest gap between human and AI reports was observed for the MIMIC-CXR abnormal cases, where for 38\% of the cases there was inter-rater agreement that the ground truth report was more accurate or descriptive of the chest radiographs. It is worth noting the difference between this task and the one performed in~\citep{tu2023generalist}, where each case was only rated by one radiologist and raters were instead asked to rank three AI-generated and one human-written report in order of preference. In that scenario, equivalence between reports was not considered.

\subsubsection*{Error correction}

In the error-correction evaluation, the expert raters are provided with (i) the chest X-ray image (a frontal view), and (ii) a radiology report for this image, consisting of the findings and impression sections. Their task is to assess the accuracy of the given radiology report by identifying errors in the report and providing suggested replacements. Before each annotation task, clinicians are asked whether the presented image is of sufficient quality for them to complete the task. They are then asked whether there is any part of the report that they do not agree with and, if so, are asked to \textbf{(i)} select the passage that they disagree with, \textbf{(ii)} select the reason for disagreement (finding I do not agree with is present; incorrect location of finding; incorrect severity of finding), \textbf{(iii)} specify whether the error is clinically significant or not, and \textbf{(iv)} provide a replacement for the selected passage. \autoref{fig:interface_error_correction} shows the labelling interface employed to perform this task. We instruct the raters beforehand that a clinically significant error is one that is potentially harmful or influences the downstream clinical decision (e.g., treatment) for the patient. We note that the raters evaluate both the ground-truth reports written by an expert and the ones generated by our model, but without the knowledge of their sources. Since the raters performing this task are different from the ones that wrote the original reports, this would also allow us to measure the degree of human errors in report-writing. Importantly, our evaluation differs from the prior work by \cite{tu2023generalist} where the original report was additionally provided as a reference and, hence, assumed accurate. 

\begin{figure}[t]
    \centering
    \includegraphics[width=\linewidth]{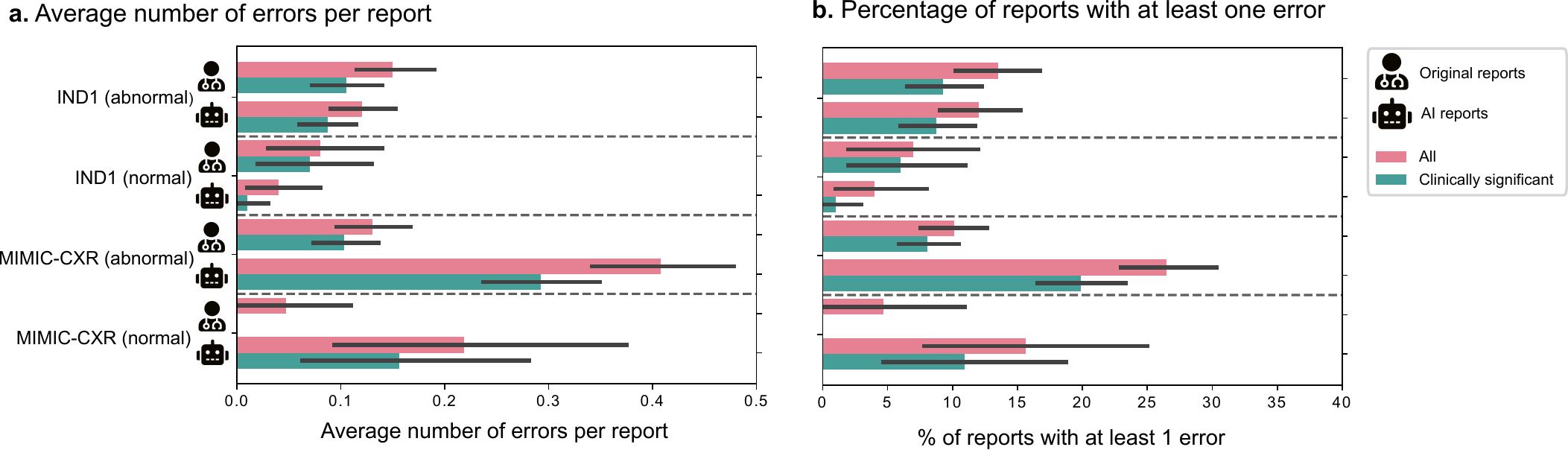}
    \caption{\small \textbf{Comparison of error correction for the AI-generated reports and the original ground truth reports.} \textbf{a.} Average number of identified (clinically significant) errors per report across both datasets for normal and abnormal cases (i.e., total number of detected errors divided by the number of all reports, including the ones without errors). \textbf{b.} Percentage of reports with at least one error for IND1 and MIMIC-CXR. Error bars correspond to 95\% confidence intervals across cases and expert assessments. Both for the normal and abnormal cases of the IND1 dataset, fewer errors are detected in AI generated reports in comparison to ground truth reports. We observe the opposite trend for the MIMIC-CXR normal and abnormal cases. It is worth noting that the ground truth reports contain a non-negligible number of clinically significant errors (except the MIMIC-CXR normal cases).}
    \label{fig:raw_errors_human_eval}
\end{figure}

Our results indicate that there is a non-negligible degree of disagreement on the original reports, especially for the abnormal cases with more than $10\%$ of the reports flagged to contain at least one error for both MIMIC-CXR and IND1 datasets as shown in \autoref{fig:raw_errors_human_eval}\textbf{.b}. We also observe that for both datasets the average number of errors per report is consistently smaller for the normal cases than the abnormal ones, likely due to the lower variability and complexity of report contents.

The relative frequency of errors between the AI system and the human experts is varied across the two datasets (\autoref{fig:raw_errors_human_eval}). For the IND1 dataset, our results show that the model makes fewer errors than the human experts; both the average number of errors per report and the proportion of reports with at least one error are lower for the model generated reports than for the original human-written ones for both normal and abnormal cases. On the other hand, the reverse is true for the MIMIC-CXR dataset. On average, 0.12 errors (0.09 of those being clinically significant) were detected in the model-generated reports among the abnormal cases in the IND1 dataset in comparison to 0.15 errors (0.11 clinically significant) in the original expert-written reports. For the normal cases in the IND1 dataset, only 0.04 errors (0.01 clinically significant) were reported on average in the generated reports, attaining a reduction from the corresponding error rates of the human experts that contained 0.09 (0.08 clinically significant) errors. For the abnormal cases of the MIMIC-CXR dataset, 0.29 clinically significant errors were reported in Flamingo-CXR reports and 0.41 total errors. The corresponding number of errors for the ground truth reports were 0.10 and 0.13, respectively. Overall, 19.9\% of Flamingo-CXR reports contained at least one clinically significant error, indicating that some cases are particularly challenging to generate reports for. This is in comparison to 8.1\% of ground truth reports on IND1 abnormal cases with at least one clinically significant error.

\begin{figure}[t!]
    \centering
    \includegraphics[width=\linewidth]{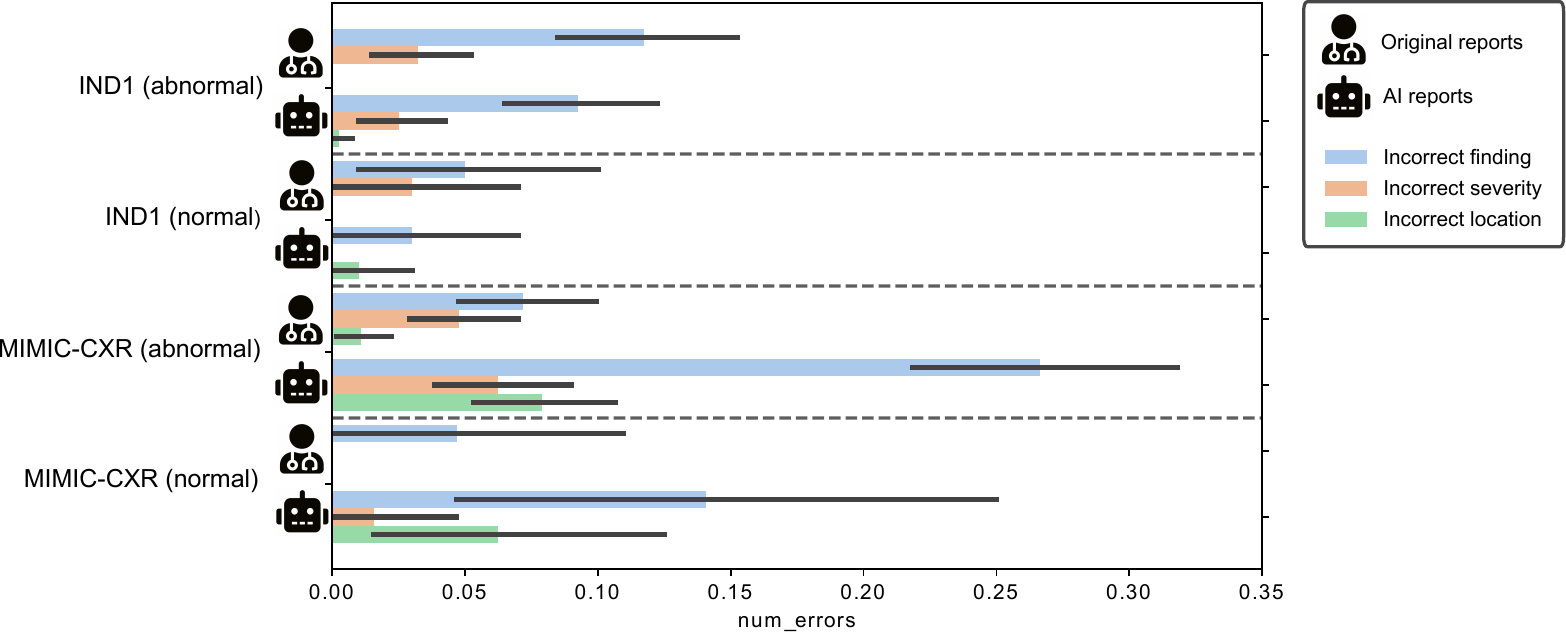}
    \caption{\small \textbf{Types of errors found in the original reports and the AI-generated reports}. During the error correction evaluation, we ask expert raters to explain the identified issues in reports based on the following taxonomy: (a) incorrect findings (this covers instances of reference to prior measurements or unavailable views), (b) incorrect severity (e.g., mild vs. severe pulmonary edema), (c) incorrect location of finding (e.g., left- vs. right-sided pleural effusion). The figure shows the distributions of these error types for the normal and abnormal cases separately in the IND1 and MIMIC-CXR datasets. The $95\%$ confidence intervals across cases are also shown. }
    \label{fig:error_taxonomy_human_eval}
\end{figure}

\begin{figure}[t!]
    \centering
    \includegraphics[width=0.95\linewidth]{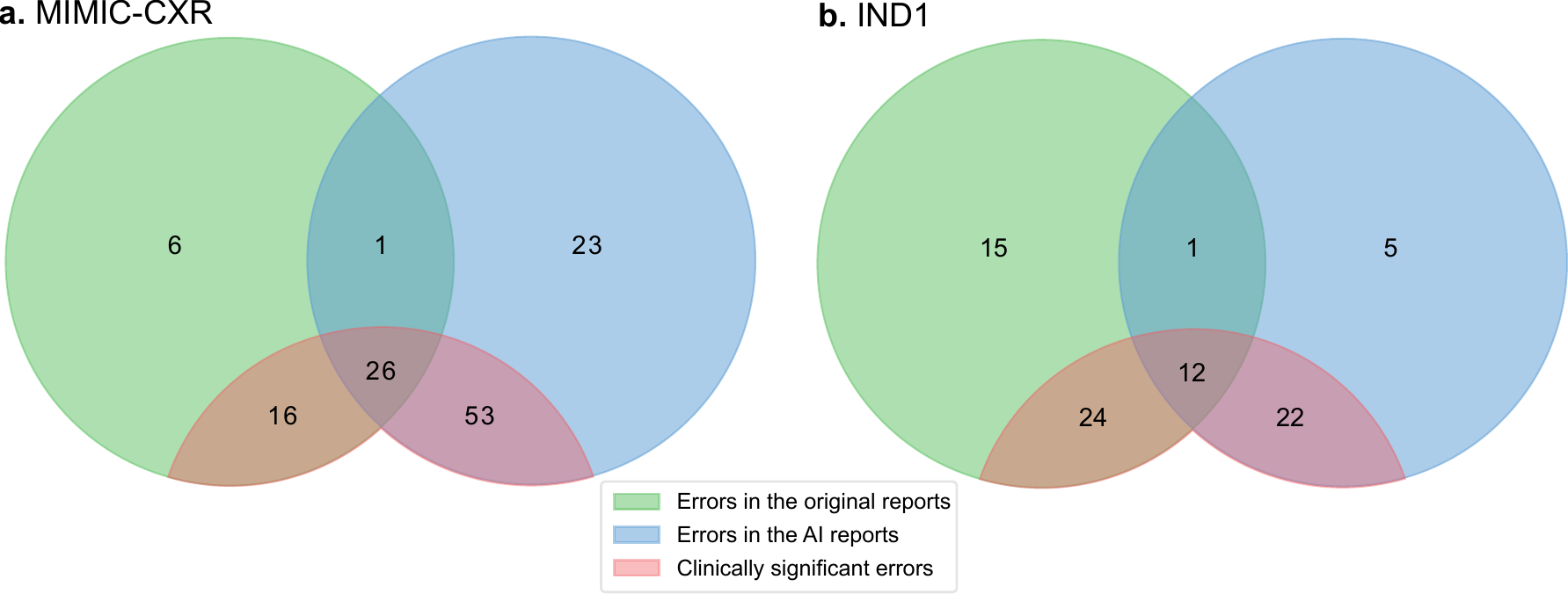}
    \caption{\small \textbf{Overlaps and disparities in cases with at least one error between the predicted reports and the original ones.} For example, the intersection between the blue and the green segments indicate the number of cases where both the AI-generated report and the ground truth were identified with errors. The red segment further indicates the cases where at least one clinically significant error is detected.
    }
    \label{fig:venn_diagram_human_eval}
\end{figure}

To compare the distributions of error types across datasets, we further show the reasons of disagreement for the edits made in reports in~\autoref{fig:error_taxonomy_human_eval}. 
For both the model-generated reports and the original ones, the most dominant category of errors is consistently the ``incorrect finding'' across the two datasets. It is worth noting that the "incorrect finding" category is less specific than the other two categories, ``incorrect severity'' and ``incorrect location'' because it also includes errors concerned with references to views or prior measurements that are not provided to the raters, but were available when the original reports were written (e.g., ``in comparison to the previous radiograph, ...''). For the AI-generated reports on the MIMIC-CXR abnormal cases, 0.27 errors on average correspond to incorrect findings, 0.08 are due to incorrect location of the finding and 0.06 due to incorrect severity. For the IND1 abnormal cases, 0.09 errors are due to incorrect findings in the predicted report and 0.03 due to incorrect severity. Overall, errors due to incorrect location of findings in the report (e.g., opacity in left vs. right lung) are more prevalent for the MIMIC-CXR abnormal cases than for the abnormal cases in IND1.

Lastly, in~\autoref{fig:venn_diagram_human_eval} we show the differences and overlap in the cases with errors between the original reports and the ones generated by our model. Large proportions of the clinically significant errors are non-overlapping (73$\%$ for MIMIC-CXR and 79$\%$ for IND1, respectively), suggesting frequent inconsistency in detected issues between the AI-reports and the original ones. Notably, in 17$\%$ and 40$\%$ of such cases, clinically significant errors were identified in the human reports but not in the corresponding AI generated reports. Some examples are provided in Appendix~\autoref{tab:clinically_significant_errors_human}, illustrating the nuanced nature of these differences. On the other hand, there is also a considerable number of instances where the AI-generated reports contain significant errors but not the original reports. Examples of such instances are provided in~\autoref{tab:clinically_significant_errors_ai}; some of these errors pertain to limited spatial reasoning and counting capabilities of VLMs. The presence of such disparities suggests potential complementarity between the AI system and the human experts in composing accurate radiology reports, motivating us to investigate the utility of CXR-Flamingo in a clinician-AI collaboration setting in~\autoref{subsec:clician_ai_collaboration}.

\subsection{Clinician-AI collaboration} \label{subsec:clician_ai_collaboration}

In this section we explore collaboration between clinicians and Flamingo-CXR as a way of leveraging the strengths of both. In the error correction task described above, our raters were given the option to suggest replacements for each of the sentences in the reports generated by our model that they disagreed with. By substituting the original statements with these replacements in the generated reports, we can produce a new set of clinician-AI reports (\autoref{fig:overview}.\textbf{b}). To evaluate the quality of these reports that result from human refinement of AI reports, we ask our expert raters to indicate their preference for clinician-AI reports relative to the corresponding original reports. Analogous to the previous setup, the raters are unaware of which report corresponds to the original ground truth and which one was initially generated by the AI model. We use the same pairwise preference interface described in~\autoref{subsec:expert_evaluation} and ensure that clinicians that previously provided corrections for the original AI reports do not perform the preference test on the same cases. We evaluate expert preferences for the same two datasets, namely IND1 and MIMIC-CXR.

\begin{figure}[t!]
    \includegraphics[width=\linewidth]{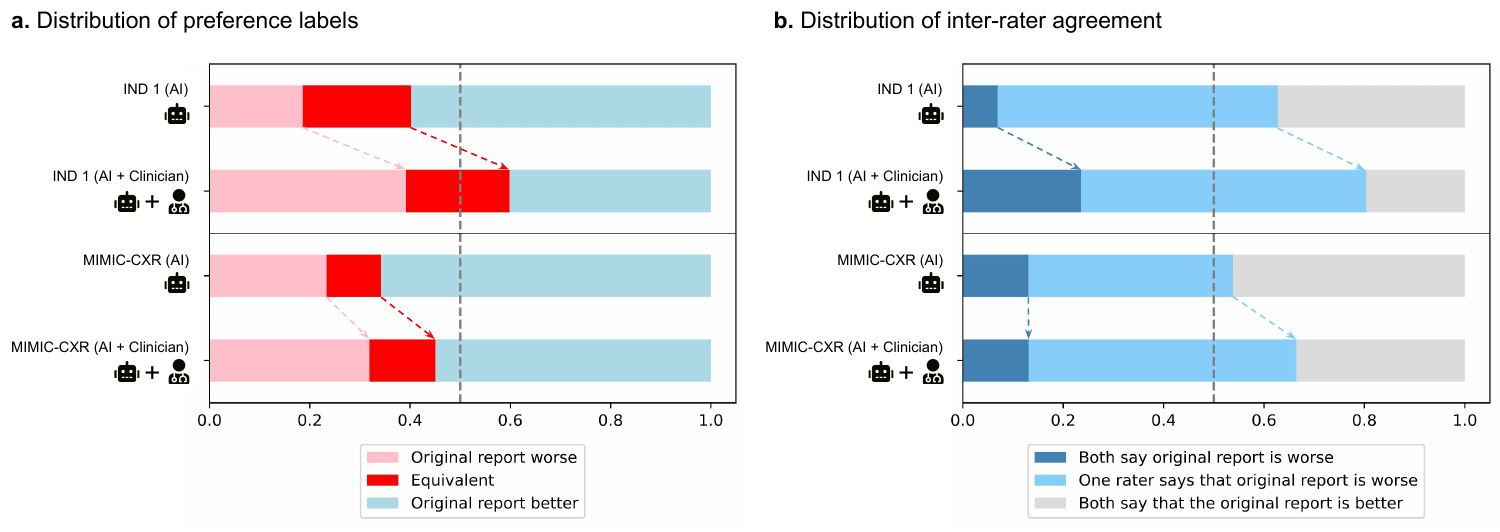}
    \caption{\small \textbf{Results of pairwise preference test for Clinician-AI collaboration}.  Preferences for reports produced from a clinician-AI collaboration are reported in comparison to preferences for original reports. We note that this figure is based on the specific subset of cases for which there was at least one edit made in the initial error correction task in~\autoref{fig:interface_error_correction}. Also, the corresponding pairwise preference scores for reports produced by Flamingo-CXR without human collaboration are also given, for the same set of reports that are edited by clinicians in the clinician-AI collaboration. \textbf{a.} Overall distribution of preferences in AI-based report vs. original report. \textbf{b.} Inter-rater agreement in their assessment (each instance is annotated by two radiologists). Colour schemes match the results shown in~\autoref{fig:clinician_AI_pairwise_preference}.}
    \label{fig:clinician_AI_pairwise_preference}
\end{figure}

For $80\%$ of IND1 cases, we find that the reports from the clinician-AI collaboration were rated as equivalent or preferred by one or more radiologist to the original ground truth report. In comparison, for reports generated by Flamingo-CXR alone without human input, the preference for model reports on the same subset of cases was $63\%$. (\autoref{fig:clinician_AI_pairwise_preference}). It is worth highlighting that we only show rater preferences for the subset of cases for which there was at least one disagreement in the original error correction task in~\autoref{fig:interface_error_correction}. We observe similar findings for MIMIC-CXR, where the reports from clinician-AI collaboration were rated as preferred or equivalent by at least one radiologist in $66\%$ of cases, in comparison to $54\%$ for reports generated by Flamingo-CXR alone.

\section{Discussion}

In this work, we present \textit{Flamingo-CXR}, a state-of-the-art AI radiology report generation system for chest radiographs built by specialising a recent vision-language foundation model \citep{alayrac2022flamingo} on this challenging task. Our model achieves competitive performance in multiple automated metrics in two clinical contexts and geographical locations, namely intensive care in the US and inpatient care delivery in India. To gauge the clinical quality and potential real-world utility of our report generation system we perform the most comprehensive expert evaluation of AI-generated reports published to date, and compare these to human-written ground-truth reports with a group of certified radiologists. This evaluation is performed both in an autonomous and an assistive AI context. Our study provides evidence
that clinicians considered AI reports equivalent or preferable to human reports in 43$\%$ and 60$\%$ of cases in the US and India datasets respectively, 
while elucidating the nuanced disparities between the two and providing meaningful directions for future enhancement. 

Prior work has repeatedly reported the shortcomings of automated ``natural language generation" metrics for assessing reports of radiology images, indicating that expert evaluation remains the gold standard for measuring both accuracy and clinical utility of report generation models~\citep{yu2022evaluating}. However, the majority of published works on the development of AI systems for this task, including recent approaches with acclaimed state-of-the-art performance, solely report automated metrics, leaving their proximity to expert accuracy and potential clinical utility unknown. Only a handful of prior works have attempted to evaluate AI systems with human experts, yet even these studies are lacking in the diversity and granularity of evaluations performed by expert radiologists. For example, \citet{tu2023generalist} employed a similar evaluation schema on the same US dataset (MIMIC-CXR) as used in this study, but assumed that the ground-truth report is correct, neglecting the inter-rater variability inherent in chest radiograph interpretation \citep{gefter2022special}. \cite{huang2023generative} recently conducted an evaluation where AI-generated reports on in-house emergency chest radiographs were compared against experts, and showed that the quality on average was only marginally inferior to that of on-site radiologists and surpassed that of teleradiology reports. However, both studies only evaluated the AI report generation model as a standalone system on a dataset acquired in an emergency department in the US; by contrast, our study considers a more diverse set-up that encompasses both autonomous and assistive scenarios for datasets from intensive care in the US as well as inpatient care delivery in India. Furthermore, our study enriches this evaluation by collecting granular information on error types (e.g., distinction between incorrect findings, location and severity), and provides fine-grained insights into how the AI system differs from human experts, which was absent in the prior works.

Human evaluation results shed more light on the aspects of our model's report quality that might inform and enable applications of the technology in future clinical workflows. For both the US- (MIMIC-CXR) and India-sourced (IND1) datasets, at least one radiologist (out of two per case) expressed preference or equivalence over the AI report to the ground truth report in more than 60$\%$ of cases. For the IND1 dataset, the number of errors are consistently lower than the ground truth reports for both normal and abnormal cases (\autoref{fig:interface_error_correction}); notably for the normal cases, the raters unanimously viewed the AI generated reports to be at least equivalent to the human reports in 77$\%$ of the cases. This strong performance on normal cases suggests potential clinical applicability in using the report generation model in the subset of such cases,
(for instance, taken alongside prior works that show AI systems to have strong accuracy in predicting whether CXRs are normal or abnormal \citep{nabulsi2021deep}),
allowing radiologist attention to be allocated to critical patients with abnormalities. On the other hand, we notice there is considerable room for improvement for the MIMIC-CXR dataset whose reports are in general more detailed and less templated than IND1. The higher error rates for the AI-generated reports in MIMIC-CXR are also consistent with the results of the pairwise preference test in that the proportion of unanimous cases is lower for the AI reports than for the ground truths (\autoref{fig:pairwise_preference}). This inter-dataset discrepancy in report quality highlights the importance of evaluation in different clinical contexts and geographic regions, which was previously not considered. The desired contents of a report are ultimately contingent on the given clinical context, and assuming access to large quantities of training data from every plausible scenario is not realistic.
Future work will consider reinforcing our system with the capability to follow user instructions \citep{singhal2023towards} so the users can control the outputs more flexibly through natural language and the capability to learn efficiently from a small quantity of data through techniques such as in-context learning \citep{moor2023medflamingo} or parameter-efficient optimisation \citep{singhal2023large}.

The complexity in evaluating the quality of radiology reports is underscored by the observed high inter-rater variability as evidenced by: (a) identified (clinically significant) errors in the ground truth reports as part of the error correction task, and (b) the variability in both human evaluation tasks in terms of preferences and disagreements with report statements. This indicates the importance of our approach to obtaining multiple readings per case, unlike prior works that have only evaluated each case once~\citep{tu2023generalist}.

In-depth analysis shows that both human and AI systems can err in different ways, hinting at potential complementary properties between the two.  Manual inspection unveils some examples where nuanced clinical errors were detected in the human reports, but not in the corresponding AI-generated reports (see~\autoref{tab:clinically_significant_errors_human}). In one case, radiologists have noted cardiomegaly in their findings, while Flamingo-CXR was able to correctly describe the cardiac silhouette as a borderline case of normality (explicated as ``top normal''). In another case, pleural thickening was also successfully distinguished from pleural effusion when the ground truth report incorrectly mentioned the former. On the other hand, AI errors often under-report findings (e.g., absence of pleural effusion when mild pleural effusion is present) and struggle with counting (see single vs. dual lead pacemaker examples in~\autoref{tab:clinically_significant_errors_ai}) and spatial reasoning. The latter is also evidenced by the high number of errors classified as referring to ``incorrect location of findings'' in the AI reports for the abnormal cases in the MIMIC-CXR dataset (\autoref{fig:error_taxonomy_human_eval}). Interestingly, this inaccuracy in object counting and spatial reasoning are also known limitations of vision-language models that have been previously corroborated by~\citet{puatruaucean2023perception} in a thorough evaluation of Flamingo~\citep{alayrac2022flamingo} and other models on a variety of non-medical benchmarks. Finally, another noteworthy difference between clinicians and our AI-system is the input information at disposal when writing the reports; Flamingo-CXR only had access to the current radiograph at a lower resolution of 1 megapixels (in contrast with the original resolution of approximately 4 megapixels) while the original radiologists additionally had access to contextual information, patient history and previous scans. Integrating such extra information into our AI system will likely enhance the reporting accuracy \citep{bannur2023learning} but requires further study. 

To our knowledge, this work is the first to evaluate the report generation approach to chest radiograph interpretation in an assistive setting. Results show that AI-generated reports with expert revisions were preferred by at least one radiologist to reports written by experts alone in 80$\%$ and 66$\%$ of cases in IND1 and MIMIC-CXR. Our proof-of-concept evaluation demonstrates the initial promise of AI report generation as an assistive system that augments the physicians' report writing process. 

These results are not without limitations. Intriguingly, the AI reports with human edits do not reach perfect preference or equivalence against the original reports. There are several possible reasons for this. Firstly, there is a baseline level of inter-rater variability both in the preference decision and the error correction process, which accounts for a certain percentage of this gap. We observed in particular, some edits made in the error correction task were of questionable quality (e.g., a whole sentence replaced with a single word e.g. "Cardiomegaly") which render the resultant reports quite unnatural despite being clinically more correct. Secondly, in certain cases, a clinician working in collaboration with AI may produce a report that is less accurate than a clinician working alone. Indeed, this is a common phenomenon observed in multiple lines of work in chest X-ray classification tasks, where collaboration often result in less accurate predictions \citep{rajpurkar2023current}. Clinician-AI collaboration typically becomes unhelpful when the experts overly rely on the AI predictions \citep{rajpurkar2020chexaid,seah2021effect} or are unduly critical about them \citep{agarwal2023combining}. Development of effective strategies for identifying when providing an AI generated report is helpful is likely needed for maximising the benefits of AI-assistance \citep{dvijotham2023enhancing}. Thirdly, while it is plausible that revising an AI-generated report may require less time than composing one from scratch, this work does not assess this explicitly yet. Quantifying the time saving aspect, however, warrants another carefully designed human study to approximate report composition time of human experts, which commonly varies between individuals and is influenced by a plethora of factors such as the clinical context, reporting style, expertise and complexity of cases. Finally, clinician-AI collaborations can take more complex forms than our design and ideally should ultimately be bi-directional and interactive, much like an experienced colleague that answers the radiologist's questions and provides high-quality feedback on their reports (e.g., flagging potential errors and missing findings). While we have witnessed initial signs of such possibilities in the recent works on interactive, multi-modal medical AI \citep{li2023llavamed,moor2023medflamingo}, there remain a considerable amount of progress to be made towards building a clinically useful writing assistant for radiology. 

\section*{Acknowledgements}
This research was funded by Google DeepMind and Google Research. We would like to thank many colleagues for useful discussions, suggestions, feedback, and advice, including Christopher Kelly, Greg S. Corrado, Jonathan Krause, Nenad Tomasev, Oriol Vinyals, Shakir Mohamed, Taylan Cemgil and Yutian Chen. 

\bibliography{references}

\newpage
\appendix

\title{Supplementary Materials}

\section{Methods} 
\label{sec:methods}

\subsection{Model} 
\label{subsec:methods_model} 
Flamingo-CXR is built by adapting a version of Flamingo \citep{alayrac2022flamingo}, a recent vision-language model to the radiology report generation task. In this section, we elaborate on the details of the task, architecture, optimisation and inference. 

\subsection{Task}Our model, Flamingo-CXR is trained to generate the findings and the impressions sections of a radiology report given a frontal chest X-ray image. This task formulation closely follows the recent studies \citep{chen2020,yan2021weakly,bannur2023learning,wang2023r2gengpt}. 

\paragraph{Architecture} 
Flamingo is a general-purpose family of transformer-based visual language models (VLMs) that take visual data as input (e.g., images) interleaved with text and produce free-form text as output. The key architectural components are (i) the language model that operates on the input text and generates the output text, (ii) the vision encoder that maps visual data into the same representation space as text input, and (iii) the connective module that integrates both modalities. The combination of the Perceiver Resampler \citep{jaegle2021perceiver} and cross-attention layers in this connective component offer an expressive way for the language model to incorporate visual information for the next-token prediction task. There are multiple versions of Flamingo at different scales, and our report generation model, Flamingo-CXR is built based on a parsimonious one with 400 million parameters. Flamingo models the likelihood of the radiology report $y$ conditioned on the input image $x$ in an auto-regressive fashion:
\begin{align}
    p(y | x) = \prod_{\ell=1}^L p(y_\ell | y_{< \ell}, x_{\leq \ell}),
    \label{eq:modeling}
\end{align}
where $y_{\ell}$ is the $\ell$-th language token of the input report, $y_{<\ell}$ is the set of preceding tokens and $p$ is parameterised by the model.

\paragraph{Optimisation} We take a version of Flamingo, pre-trained on a large set of interleaved text-image data, and finetune it on the specific task of radiology report generation by minimizing a weighted sum of the expected negative log-likelihoods of report given the chest radiograph over both MIMIC-CXR (US) and IND1 (India) datasets:
\begin{equation}
     \lambda_{\text{US}} \cdot \mathbb{E}_{(x, y)\sim \mathcal{D}_{\text{US}}} \left[ -\sum_{\ell=1}^L w(x,y) \cdot \log p(y_\ell | y_{< \ell}, x_{\leq \ell})\right] +  \lambda_{\text{India}} \cdot \mathbb{E}_{(x, y)\sim \mathcal{D}_{\text{India}}} \left[ -\sum_{\ell=1}^L w(x,y) \cdot \log p(y_\ell | y_{< \ell}, x_{\leq \ell})\right], 
\end{equation}
where $\mathcal{D}_{\text{US}}$ and $\mathcal{D}_{\text{India}}$ denote the MIMIC-CXR and IND1 datasets respectively, $\lambda_m$ are the data-specific coefficients that are tuned to maximise the benefits of jointly training on both datasets, and lastly
$w(x, y)$ is a re-weighting function that changes the amount of penalty depending on whether the example $(x, y)$ contains any thoracic abnormalities. Specifically, we employ \textit{importance weighting} \citep{horvitz1952generalization} here and define $w(x, y)$ to output the inverse of the proportion of healthy cases in the corresponding dataset (if the given example is so) or otherwise that of abnormal cases. This ensures that the model is equally penalised to compose inaccurate reports across the healthy and the abnormal cases; this is particularly important for the IND1 dataset where the healthy cases account for more than 90$\%$ of the training data. 

To further enhance the reporting accuracy on abnormal cases, we augment the above training objective with an auxiliary classification loss for abnormality classification. To this end, we applied a published labelling software, CheXpert \citep{irvin2019chexpert} to extract the presence of multiple thoracic conditions from the training reports, derived binary abnormality labels (1 if any of the conditions is present or else 0), and used them to compute this auxiliary classification loss. We found the addition of this abnormality classification task to be helpful in improving the sensitivity of the generated reports across these conditions. 

We optimize parameters using AdamW~\citep{loshchilov2018fixing} with initial learning rate of $10^{-3}$ and $\beta = [0.9, 0.999]$ with batch size of 16 examples and train for 150,000 steps. The best checkpoint was selected based on the overall CIDEr-d score on the validation set. We freeze the language component and only update the parameters in the vision encoder and the connective component (perceiver resampler and cross-attention layers) as our initial experiments showed updating the language part resulted in overfitting and finetuning the rest of the architecture was important for adapting to the unfamiliar medical domain not represented in the pre-training datasets.

\paragraph{Inference} \label{sec:inference}
Once Flamingo is trained, we use it generate the radiology reports on the test chest radiographs with two decoding strategies: \textit{beam search} with the width size set to 3 and \textit{nucleus sampling} with $p=0.9$ \citep{holtzman2019curious}. We employed the former deterministic decoding method by default, and the generated reports are used in calculating of reported NLG and clinical metrics in \autoref{tab:table_mimic_appendix} and \autoref{tab:table_apollo} as well as in the subsequent expert evaluation. On the other hand, we also used the latter stochastic decoding method when we needed to generate multiple reports. For example, to plot the ROC curves in \autoref{fig:apollo_overall} and \autoref{fig:apollo_roc_per_condition} for measuring the disease classification accuracy of reports, we used the nucleus sampling to generate 250 candidate reports, derived the condition labels from each with the CheXpert labeller and aggregated them to compute the per-condition probability.

\subsection{Datasets and pre-processing} \label{sec:methods_datasets}

\paragraph{IND1} A de-identified dataset of 263,021  frontal chest radiographs (digital and scanned) with reports obtained from five regional centers across a large hospital group in India (Bangalore, Bhubaneswar, Chennai, Hyderabad, and New Delhi) between November 2010 and January 2018 \citep{ahn2022association}. We use the same training, validation and test split as done in \citep{nabulsi2021deep}. Thus, a total of 250,066 samples are used for training, 4,960 samples for validation, and 7,995 samples for testing of Flamingo-CXR. Furthermore, a small subset of 2306 cases are annotated with varying numbers of binary labels (0: absent, 1: present) for 6 thoracic conditions (Cardiomegaly, Pleural Effusion, Lung Opacity, Edema, Enlarged Cardiomediastinum and Fracture) obtained from a pool of 18 certified radiologists in the US. The consensus labels are derived by calculating the majority vote, and used as the reference labels for evaluation of report quality in classification accuracy (e.g., ROC curves in \autoref{fig:apollo_roc_per_condition} and F1 scores in \autoref{tab:table_apollo}). 

\paragraph{MIMIC-CXR} As the largest public dataset to date, MIMIC-CXR \citep{johnson2019mimic} contains 377,110 images and 227,835 reports. In our experiments, we use the official split provided by the dataset resulting in 222,758 training examples, 1,808 validation examples and 3,269 test examples. For the reports, we remove redundant whitespaces (i.e., line breaks, etc). We only use frontal view scans (AP and PA views) and discard samples where only lateral views are provided. Following \citep{bannur2023learning}, we only keep the FINDINGS and IMPRESSION sections of reports and filter out cases that does not contain an IMPRESSION section. 

Lastly, more than 50$\%$ of the examples in MIMIC-CXR contain prior scans \cite{bannur2023learning} and the corresponding reports often describe findings in reference to these measurements (see the highlighted sentence in the left column of \autoref{tab:examples_with_prior_refs} for an example). Consequently, as also reported in a recent work by \cite{pmlr-v193-ramesh22a}, naively training on the entirety of the MIMIC-CXR data leads to a model that generates reports with hallucinated references to non-existent prior reports (see the right column; note that the model only has access to the current radiograph). To ameliorate this issue, we remove all the training examples with references to prior studies (see the middle column for an example of the improved prediction as a result). However, we still report the evaluation metrics on all the test examples for a fair comparison with the previous studies. The combination of all the above pre-processing and filtering steps result in 90,968 training, 688 validation and 1931 test examples.

\begin{table}[t!]
    \centering
    \caption{\small \textbf{Examples of references to the prior measurements both in the original and the predicted reports from Flamingo-CXR trained with/without the proposed pre-processing.} The descriptions in relation to the prior data are highlighted. }
    \label{tab:examples_with_prior_refs}
    \scriptsize
    \begin{tabular}{|p{0.3\linewidth}||p{0.3\linewidth}||p{0.3\linewidth}|}
        \hline
        \textbf{Original} & \textbf{Prediction (with pre-processing)} & \textbf{Prediction (without pre-processing)}   \\ \hline
        FINDINGS: \newline
        AP view of the chest. The patient has been intubated \colorbox{yellow}{since prior exam.} Right PICC is seen with tip at the upper SVC. Relatively low lung volumes are seen. The lungs however remain clear without consolidation, effusion or pulmonary vascular congestion .  Cardiac silhouette appears moderately enlarged, likely accentuated due to low lung volumes and AP technique. 
       & FINDINGS:\newline
       Single AP upright portable view of the chest was obtained. A right-sided PICC is seen, terminating at the cavoatrial junction/right atrium. The cardiac silhouette appears mildly enlarged. Mediastinal contours are stable. There is no overt pulmonary edema. No large pleural effusion is seen, although a trace right pleural effusion would be difficult to exclude. 
       & FINDINGS: \newline
       \colorbox{yellow}{As compared to the previous radiograph,} the patient has been intubated. The tip of the endotracheal tube projects approximately 3.5 cm above the carina. There is no evidence of complications, notably no pneumothorax. The other monitoring and support devices are constant, with the exception of the nasogastric tube \colorbox{yellow}{that has been newly inserted.} The course of the tube is unremarkable, the tip of the tube projects over the middle parts of the stomach. \colorbox{yellow}{Unchanged} evidence of bilateral parenchymal opacities and of bilateral pleural effusions. \colorbox{yellow}{Unchanged} borderline size of the cardiac silhouette.
       
       \\ \hline
 \end{tabular}
\end{table}

\paragraph{Image processing}
All images in both datasets are resized to $320\times320$ while preserving the original aspect ratio, padded if needed, and normalized to zero mean and unit standard deviation. Color jitter and resize/crop transformations are applied as data-augmentation during the training of Flamingo-CXR.

\subsection{Expert evaluation} \label{sec:human_eval_details}
\paragraph{Annotators}
We recruited a group of 16 certified radiologists in India. All raters performed the  required CITI training before performing the evaluation tasks on the MIMIC-CXR dataset. We should highlight that radiologists that provided annotations for the first phase of error correction or preference test tasks were excluded from the human-AI collaboration evaluation to avoid annotation bias. Prior to the large scale evaluation, we validated the labelling interface with an expert to ensure that instructions were clear and opt-out options were available where essential. 

\paragraph{Sample selection}
We randomly select a fixed number of normal and abnormal cases from the IND1 and MIMIC-CXR dataset. To ensure good coverage of different abnormalities the set of abnormal cases reviewed by radiologists was larger than the one for normal cases. It is also worth noting that the same set of cases was annotated in both error correction and pairwise preference tasks. For the \textit{MIMIC-CXR} dataset, we include cases annotated in the human evaluation of the prior work by \cite{tu2023generalist} that survived the filtering stage described in~\autoref{sec:methods_datasets}.

\begin{figure}[bp!]
    \centering
    \includegraphics[width=\linewidth]{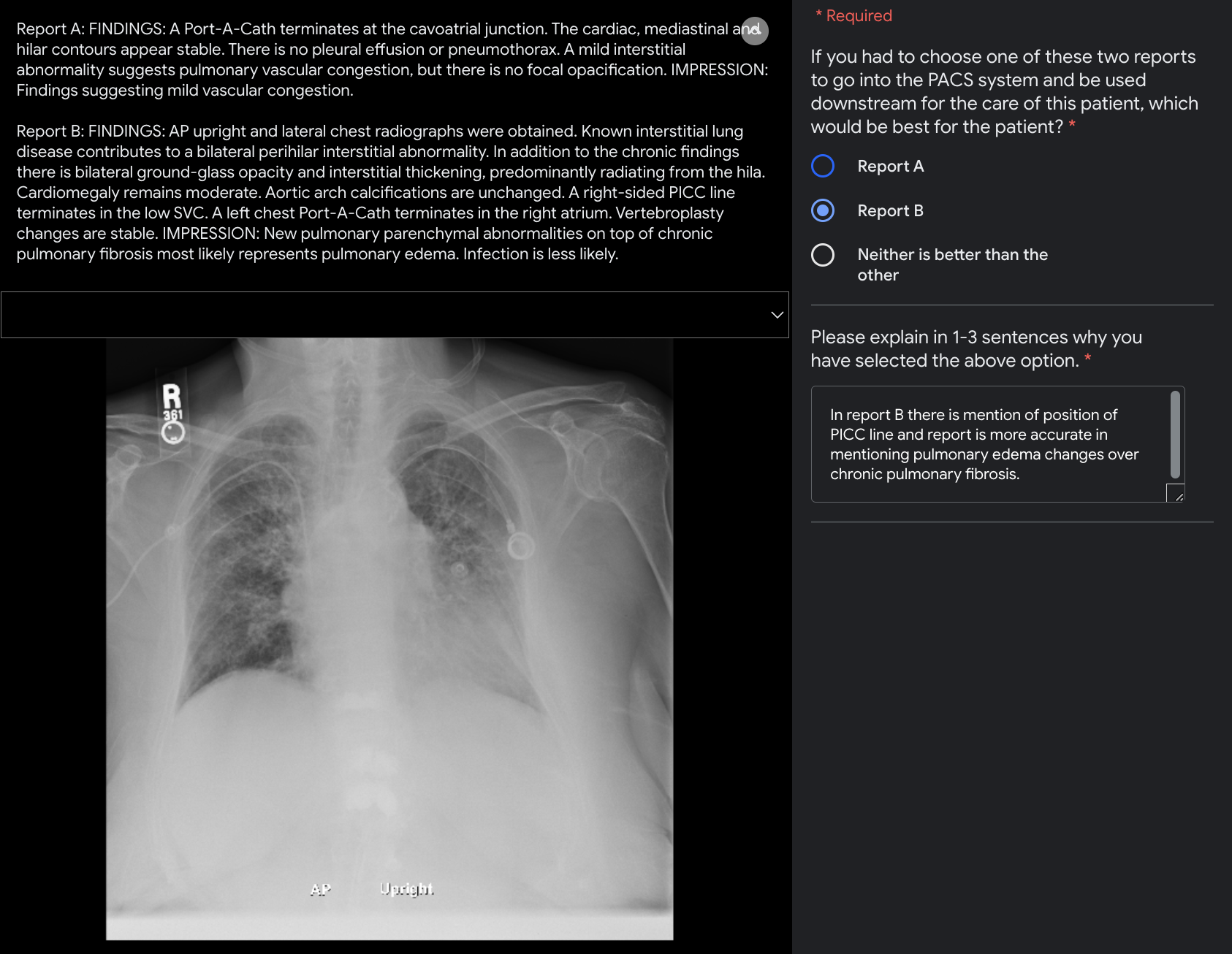}
    \caption{\small \textbf{Labelling interface for pairwise preference test}. Raters are provided with (i) a frontal view (PA or AP) in the original resolution, (ii) a radiology report generated by our AI system and (iii) the original report written by a radiologist, and are asked to provide their preference. For each case, the raters are unaware of which report is the ground-truth and which one is generated by our model, and are requested to describe their preference out of three options; report A, report B, or equivalence between the two (i.e., “neither is better than the other”). The interface allows the raters to zoom in and out on the image as needed. They are additionally asked to provide an explanation for their choice.}
    \label{fig:interface_pairwise_preference}
\end{figure}

\begin{figure}[bp!]
    \centering
    \includegraphics[width=\linewidth]{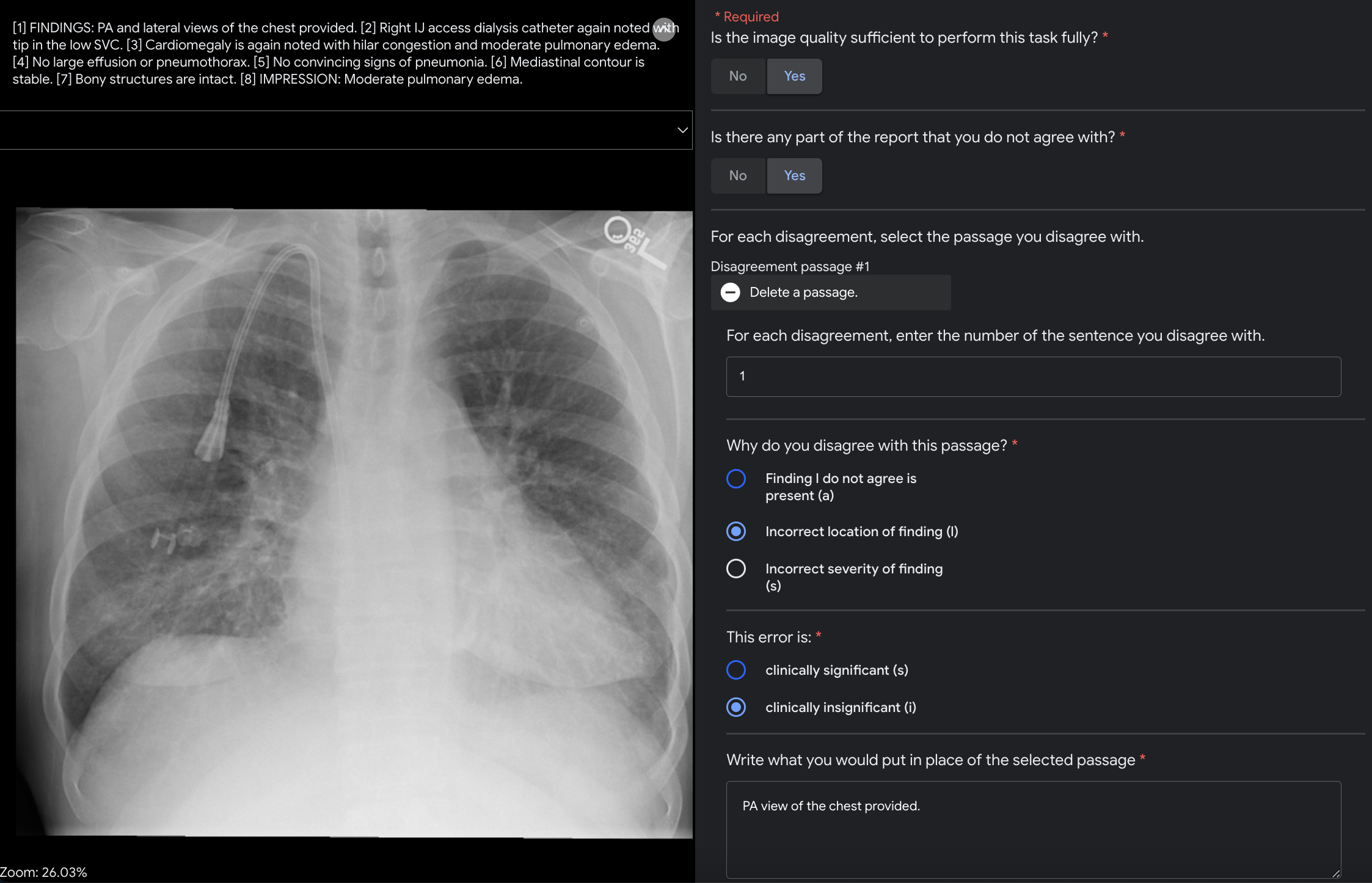}
    \caption{\small \textbf{Labelling interface for error correction task}. Raters are provided with (i) the chest X-ray image (a frontal view) and (ii) a radiology report for this image, consisting of the findings and impression sections. Their task is to assess the accuracy of the given radiology report by identifying errors in the report and correcting them. Before each annotation task, clinicians are asked whether the presented image is of sufficient quality for them to complete the task. They are then asked whether there is any part of the report that they do not agree with and, if so, are asked to (a) select the passage that they disagree with, (b) select the reason for disagreement (finding I do not agree with is present; incorrect location of finding; incorrect severity of finding), (c) specify whether the error is clinically significant or not, and (d) provide a replacement for the selected passage. }
    \label{fig:interface_error_correction}
\end{figure}

\paragraph{Annotation interface} 
We employ an internal platform for data collection to perform our expert evaluation. \autoref{fig:interface_pairwise_preference} and \autoref{fig:interface_error_correction} illustrate the labelling interfaces used by our raters to perform the pairwise preference and error correction tasks. All data were stored in the Digital Imaging and Communications in Medicine (DICOM) format and de-identified prior to transfer to the external radiologists for annotation. Experts were asked to confirm whether the image provided to them for each task was of sufficient quality for them to complete the task.

\section{Additional results}
\subsection{Automated evaluation using NLG and clinical metrics}
\begin{figure}[bp!]
    \centering
    \includegraphics[width=\linewidth]{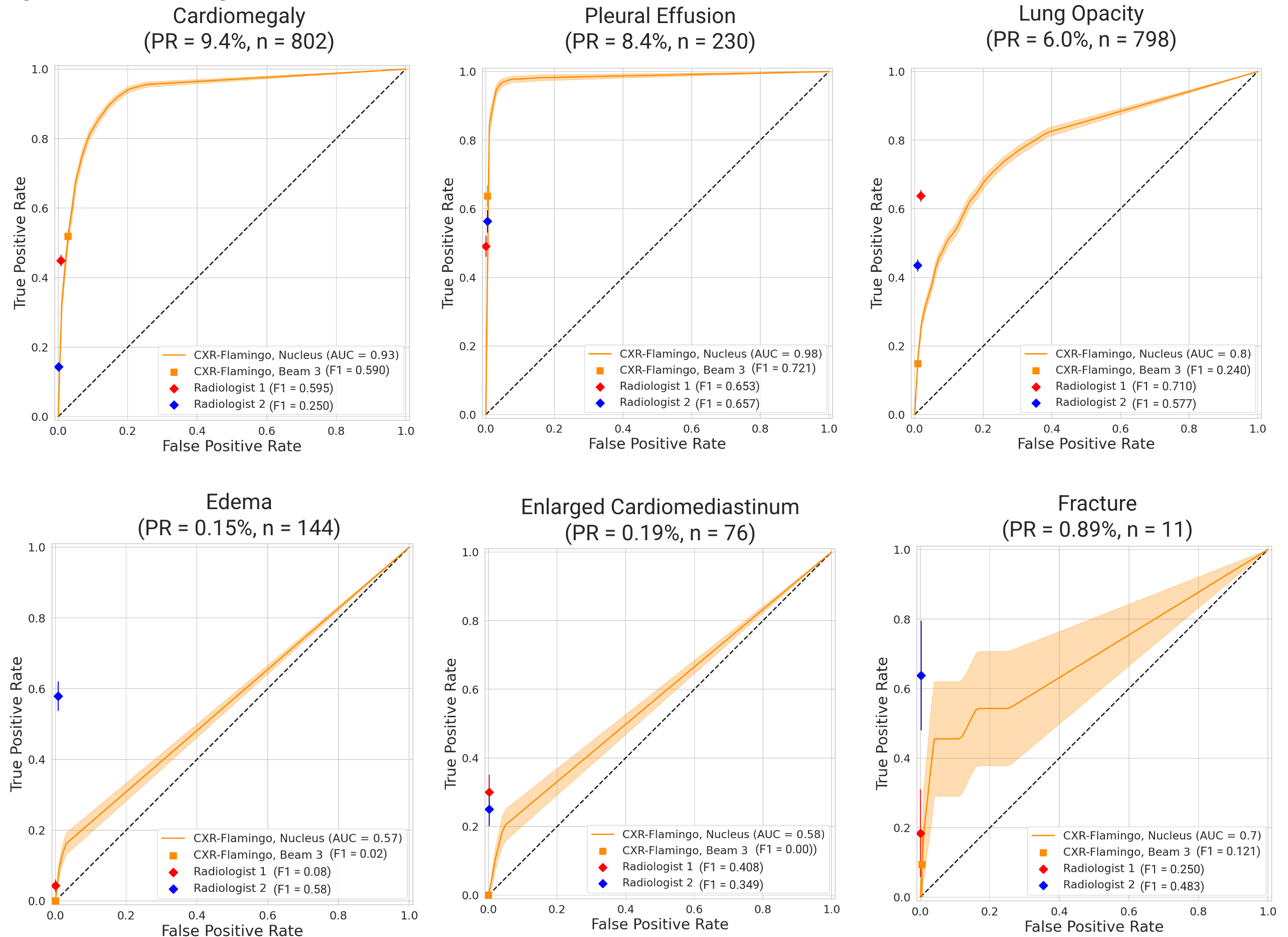}
    \caption{\small \textbf{Detection accuracy per condition on the IND1 dataset.} The receiver operating characteristic (ROC) curve of the Flamingo-CXR report generation model, shown along with the true positive rate (TPR) and false positive rate (FPR) pairs for two certified radiologists are shown for 6 conditions for which the expert labels were collected. The operating point of our model with the default inference scheme (Beam 3) is also shown.}
    \label{fig:apollo_roc_per_condition}
\end{figure}

Here we provide more detailed results on the automated evaluation of CXR-Flamingo on both IND1 and MIMIC-CXR datasets that are precluded from the main text. \autoref{tab:table_mimic_appendix} shows a comparison of our model with other prior methods in the NLG metrics as well as the clinical metrics. \autoref{tab:table_apollo} shows the same set of metrics on the IND1 test set. \autoref{fig:apollo_roc_per_condition} expands on \autoref{fig:apollo_overall} in the main paper by showing the ROC curves for six clinical conditions separately (namely, Cardiomegaly, Pleural Effusion, Lung Opacity, Edema, Enlarged Cardiomediastinum and Fracture).

\begin{table}[bp!]
    \centering
    \caption{\small \textbf{Comparison of automatic report generation metrics on the MIMIC-CXR dataset.} The column `Sections' indicates which sections of the radiology reports are generated by the respective models; `F' indicates  FINDINGS and `I` indicates IMPRESSIONS sections. Note that the metrics are retrieved from the corresponding publications. For all metrics, the higher (the bluer) the better, and the best results are shown in bold. }
    \label{tab:table_mimic_appendix}
    \scriptsize
    \centering
    \resizebox{\columnwidth}{!}{%
    \begin{tabular}{|*{8}{c|}}
        \hhline{~|~*{6}{|-}}
        \multicolumn{1}{c}{} & \multicolumn{1}{c|}{} & \multicolumn{3}{c|}{\textbf{NLG Metrics}} &\multicolumn{3}{c|}{\textbf{Clinical Metrics}}  \\\hline
        \textbf{Model} & \textbf{Sections} & \textbf{CIDEr} & \textbf{BLEU4} & \textbf{Rouge}  & \textbf{F1 (all)} & \textbf{F1 (top 5)} & \textbf{Radgraph}\\ \hline
        CXR-RePaiR \citep{Krishnan2021} & F & - & \gradientblu{0.021} & \gradientrou{0.143} & \gradientall{0.281} & - & \gradientrad{0.091}   \\ \hline 
        $\mathcal{M}^2$ Transformer \citep{miura2021} & F & \textbf{\gradientcdr{0.509}} & \gradientblu{0.114} &  - & - & \gradienttop{0.567} & \gradientrad{0.220}   \\ \hline
        RGRG \citep{tanida2023interactive} & F & \gradientcdr{0.495} & \gradientblu{0.126} & \gradientrou{0.264} & \gradientall{0.447} & \gradienttop{0.547} & -   \\ \hline
        METransformer \citep{wang2023metransformer} & F & \gradientcdr{0.362} & \gradientblu{0.124} & \gradientrou{0.291} & \gradientall{0.311} & - & -   \\ \hline
        Med-PaLM-M, 12B \citep{tu2023generalist}& F &\gradientcdr{0.234} & \gradientblu{0.104} & \gradientrou{0.262} & \gradientall{0.514} & \gradienttop{0.565} & \textbf{\gradientrad{0.252}}   \\ \hline \hline
        R2Gen \citep{chen2020} &  F + I & - & \gradientblu{0.103} & \gradientrou{0.277} & \gradientall{0.228} & \gradienttop{0.346} & \gradientrad{0.134}   \\ \hline
         WCT \citep{yan2021weakly} & F + I & - & \textbf{\gradientblu{0.144}} & \gradientrou{0.274} & \gradientall{0.294} & - & \gradientrad{0.143}   \\ \hline
         CvT-21DistillGPT2 \citep{NICOLSON2023} & F + I & \gradientcdr{0.361} & \gradientblu{0.124} & \gradientrou{0.285} & \gradientall{0.384} & - & \gradientrad{0.154}   \\ \hline
        BioVil-T \citep{bannur2023learning} & F + I & - & \gradientblu{0.092} & \gradientrou{0.296} & \gradientall{0.317} & - & -   \\ \hline
        R2GenGPT \citep{wang2023r2gengpt} & F + I & \gradientcdr{0.269} & \gradientblu{0.134} & \textbf{\gradientrou{0.297}} & \gradientall{0.389} & - & -   \\ \hline 
        \cellcolor{blue!15} Flamingo-CXR (Ours) & F + I & \gradientcdr{0.138} & \gradientblu{0.101} & \textbf{\gradientrou{0.297}} & \textbf{\gradientall{0.519}} & \textbf{\gradienttop{0.580}} & \gradientrad{0.205}   \\ \hline
    \end{tabular}
    }
\end{table}

\begin{table}[bp!]
    \centering
    \caption{\small \textbf{Automated report generation metrics on the IND1 dataset.} We note that there are no published report generation metrics due to the private nature of the dataset. The disease classification accuracy (F1 scores) are also computed for two radiologists. }
    \label{tab:table_apollo}
    \scriptsize
    \begin{tabular}{|*{7}{c|}}
        \hhline{~*{6}{-}}
        \multicolumn{1}{c|}{} & \multicolumn{3}{c|}{\textbf{NLG Metrics}} &\multicolumn{3}{c|}{\textbf{Clinical Metrics}}  \\\hline
        \textbf{Model} & \textbf{CIDEr} & \textbf{BLEU4} & \textbf{Rouge-L}  & \textbf{F1 (all)} & \textbf{F1 (top 3)} & \textbf{Radgraph}\\ \hline
        Flamingo-CXR (Ours) & 5.158 & 0.724 & 0.851 & 0.463 & 0.512 & 0.805  \\ \hline
       Radiologist 1  & - & - & - & 0.621 & 0.657 &  -  \\ \hline
       Radiologist 2  & - & - & - & 0.467 & 0.476 &  -  \\ \hline
 \end{tabular}
\end{table}

\subsection{Additional qualitative results} 
Here we provide some examples of both generated and ground-truth reports with large deviations. \autoref{tab:clinically_significant_errors_human} shows some examples where clinically significant errors were documented in the human reports but not in the corresponding AI generated reports, illustrating the nuanced nature of these differences. On the other hand, there are also numerous instances where the AI-generated reports contain significant errors but not the original ones, some of which are given in \autoref{tab:clinically_significant_errors_ai}.

\begin{table}[bp!]
    \centering
    \caption{\small \textbf{Examples of cases with clinically significant errors found in the ground truth radiology report, but not in the predicted Flamingo-CXR report.}}
    \label{tab:clinically_significant_errors_human}
    \scriptsize
    \begin{tabular}{|p{0.33\linewidth}||p{0.33\linewidth}|p{0.12\linewidth}|p{0.12\linewidth}|}
        \hline
        \textbf{Predicted report} & \textbf{Radiologist report} & \textbf{Replacement} & \textbf{Disagreement reason}\\ \hline
       FINDINGS: Single supine AP portable view of the chest was obtained. No focal consolidation, pleural effusion, or evidence of pneumothorax is seen. \colorbox{green}{The cardiac silhouette is top normal.} The aorta is calcified and tortuous. No displaced fracture is seen. \newline IMPRESSION: No acute cardiopulmonary process. & FINDINGS: The lungs are low in volume but otehrwise clear. Left hemidiaphragm is somewhat obscured in its lateral-most component, though this could be projectional. The left lung base is poorly imaged. There is no definite pleural effusion or pneumothorax. \colorbox{pink}{Stable marked cardiomegaly is noted.}  \newline IMPRESSION: Questionable opacity in left base. When the patient's clinical status improves, repeat evaluation by PA and lateral chest radiograph is recommended to exclude a pleural effusion or left basilar parenchymal process. & Increased cardiac silhouette is likely due to position and technical region and not true cardiomegaly. & Finding I do not agree is present. \\ \hline
       FINDINGS: Frontal and lateral radiographs of the chest demonstrate stable post-radiation paramediastinal fibrosis and scarring in the right upper lobe. There is a \colorbox{green}{small right-sided pleural effusion} with adjacent atelectasis. The cardiomediastinal and hilar contours are unchanged. There is no pneumothorax. \newline IMPRESSION: Small right-sided pleural effusion with adjacent atelectasis.  & FINDINGS: An extensive right hilar lung mass is associated with radiation fibrosis, better delineated on CT \_\_\_. An additional component of postobstructive pneumonia may be present. Retrocardiac opacity, left pleural effusion, and left plueral thickening are also new. No pneumothorax is present. \newline IMPRESSION: 1. Large right hilar lung mass and radiation fibrosis. Additional post-obstructive pneumonia in the right upper and lower lobes is possible but hard to delineate. 2. New left retrocardiac opacity, small left effusion, \colorbox{pink}{and pleural thickening}. Findings were discussed with \_\_\_, RN, via telephone at \_\_\_ and again with Dr \_\_\_ at \_\_\_. & In addition, right pleural effusion versus thickening. & Incorrect severity of finding. \\ \hline
       FINDINGS: Severe cardiomegaly is re- demonstrated. The mediastinal and hilar contours are unchanged. There is \colorbox{green}{mild pulmonary edema}, worse in the interval. No focal consolidation, pleural effusion or pneumothorax is present. There are no acute osseous abnormalities. \newline IMPRESSION: Severe cardiomegaly with mild pulmonary edema, worse in the interval. & FINDINGS: \colorbox{pink}{The lungs are well expanded} \colorbox{pink}{and clear.} Area of increase density overlying the right hilum with a sharp lower margin is of unclear clinical significance. Severe cardiomegaly is reidentified. The hilar contours are unremarkable. There is no pleural effusion or pneumothorax. \newline IMPRESSION: 1. Area of increase density overlying the right hilum with a sharp lower margin is of unclear clinical significance. Chest CT is recommended for further assessment. 2. Severe cardiomegaly, unchanged. The impression was entered as an urgently flagged wet read on the ED dashboard by Dr \_\_\_ on \_\_\_ at 9:05 am after discussion with the attending as the patient was still in the ED. & I would have written - mild pulmonary vascular congestion. & Finding I do not agree is present. \\ \hline
 \end{tabular}
\end{table}

\begin{table}[bp!]
    \centering
    \caption{\small \textbf{Examples of cases with clinically significant errors found in the Flamingo-CXR report, but not in the original ground-truth report.}}
    \label{tab:clinically_significant_errors_ai}
    \scriptsize
    \begin{tabular}{|p{0.33\linewidth}||p{0.33\linewidth}|p{0.12\linewidth}|p{0.12\linewidth}|}
        \hline
        \textbf{Predicted report} & \textbf{Radiologist report} & \textbf{Replacement} & \textbf{Disagreement reason}\\ \hline
       FINDINGS: \colorbox{pink}{Dual lead left-sided pacemaker} \colorbox{pink}{is seen with leads extending the expected} \colorbox{pink}{positions of the right atrium and right} \colorbox{pink}{ventricle}. No focal consolidation is seen. No pleural effusion or pneumothorax is seen. The cardiac and mediastinal silhouettes are stable. \newline IMPRESSION: No acute cardiopulmonary process. & FINDINGS: \colorbox{green}{Single lead pacemaker in situ} \colorbox{green}{with the lead tip in the right ventricle}. No cardiomegaly. No features of cardiac decompensation. Prominent pulmonary arteries suggesting pulmonary arterial hypertension. No pleural effusion. Consolidation in the left lower lobe. \newline IMPRESSION: Suspected left lower lobe pneumonia. & Single chamber pace maker with tip is in right ventricle. & Finding I do not agree is present. \\ \hline
       FINDINGS: Lung volumes are low. Heart size is accentuated as a result, and appears mild to moderately enlarged. \colorbox{pink}{The aorta is tortuous and diffusely} \colorbox{pink}{calcified.} There is crowding of the bronchovascular structures with mild pulmonary edema noted. No large pleural effusion or pneumothorax is seen. Patchy opacities in the lung bases likely reflect areas of atelectasis. \newline IMPRESSION: Mild pulmonary edema. Low lung volumes. & FINDINGS: Persistent pulmonary opacities, vascular engorgement and septal lines reflect mild pulmonary edema. Small left pleural effusion cannot be excluded. Low lung volumes limit assessment of cardiomediastinal silhouette though the cardiac size appears mildly enlarged. \newline IMPRESSION: Unchanged mild pulmonary edema with likely small left pleural effusion. & Only aortic knuckle calcification is present. & Incorrect severity of finding.
       \\ \hline
       FINDINGS: Frontal and lateral chest radiographs were obtained. A right-sided Port-A-Cath terminates in the lower SVC. The lungs are fully expanded and clear. The cardiomediastinal silhouette, hilar contours, and pleural surfaces are normal. \colorbox{pink}{There is no pleural effusion} or pneumothorax. \newline IMPRESSION: No focal consolidation to suggest pneumonia. & FINDINGS: A right-sided Port-A-Cath tip sits in the lower portion of the SVC. The heart and mediastinal contours are within normal limits. The lungs are largely clear with only minimal atelectasis in the right base in accordance \colorbox{green}{with a small right pleural effusion}. There is no pneumothorax. \newline IMPRESSION: Small right pleural effusion with associated atelectasis; no pneumothorax. & There is mild right sided pleural effusion. & Finding I do not agree is present.
       \\ \hline
       FINDINGS: The lungs are well expanded and clear. The hila and pulmonary vasculature are normal. No pleural effusions or pneumothorax. The cardiomediastinal silhouette is normal. \colorbox{pink}{A left pectoral pacemaker is seen with} \colorbox{pink}{transvenous leads in the right atrium} \colorbox{pink}{and right ventricle.} \newline IMPRESSION: No acute cardiopulmonary process. & FINDINGS: The lungs appear clear. \colorbox{green}{A pacemaker is seen projecting over} \colorbox{green}{the left chest with a wire appropriately} \colorbox{green}{placed in the right atrium}. The cardiomediastinal silhouette, hilar contours, and pleural structures are normal. No pneumothorax or pleural effusion. Other than the pacemaker, no radio-opaque metallic foreign object is identified in chest radiograph. \newline IMPRESSION: 1. Pacemaker seen projecting over the left chest with a wire appropriately placed in the right atrium. Other than the pacemaker, no radiopaque metallic foreign object is identified. 2. No acute cardiopulmonary process. & Single chamber pace maker with lead in right atrium. & Incorrect location of finding.
       \\ \hline
 \end{tabular}
\end{table}

\end{document}